\documentclass{article}

\usepackage{tikz}
\usepackage{mathtools}
\usepackage{colortbl}
\usepackage{amsfonts}
\usetikzlibrary{calc,positioning}

\newcommand{\norm}[1]{\left\lVert#1\right\rVert}

\definecolor{c1}{RGB}{0,0,255}
\definecolor{c2}{RGB}{255,0,0}
\definecolor{c3}{RGB}{0,255,0}
\definecolor{c4}{RGB}{0,0,44}
\definecolor{c5}{RGB}{255,26,185}
\definecolor{c6}{RGB}{255,211,0}
\definecolor{c7}{RGB}{0,88,0}
\definecolor{c8}{RGB}{132,132,255}
\definecolor{c9}{RGB}{158,79,70}
\definecolor{c10}{RGB}{0,255,193}
\definecolor{c11}{RGB}{0,132,149}
\definecolor{c12}{RGB}{0,0,123}
\definecolor{c13}{RGB}{149,211,79}
\definecolor{c14}{RGB}{246,158,220}
\definecolor{c15}{RGB}{211,18,255}
\definecolor{c21}{RGB}{246,132,9}

\begin{document}

\title{First Passage Value}
\author{Cenk Oguz Saglam and Katie Byl}
\maketitle

\begin{abstract}

For many stochastic dynamic systems, the Mean First Passage Time (MFPT) is a useful concept, which gives expected time before a state of interest. This work is an extension of MFPT in several ways. (1) We show that for some systems the system-wide MFPT, calculated using the second largest eigenvalue only, captures most of the fundamental dynamics, even for quite complex, high-dimensional systems. (2) We generalize MFPT to Mean First Passage Value (MFPV), which gives a more general value of interest, e.g., energy expenditure, distance, or time. (3) We provide bounds on First Passage Value (FPV) for a given confidence level. At the heart of this work, we emphasize that for our goals, many hybrid systems can be approximated as Markov Decision Processes. So, many systems can be controlled effectively using this framework. However, our framework is particularly useful for metastable systems. Such systems exhibit interesting long-living behaviors from which they are guaranteed to inevitably escape (e.g., eventually arriving at a distinct failure or success state). Our goal is then either minimizing or maximizing the value until escape, depending on the application.

\end{abstract}

\section{Introduction}

First Passage Time, aka First Hitting Time, gives survival duration by answering ``how long it will take on average before a specific event (or set of events) occurs''. In discrete time models, one usually calculates the expected number of discrete time steps of survival, which corresponds to Mean First Passage Time (MFPT). While different initial conditions (states) generally result in different MFPTs, for some systems a scalar called system-wide MFPT is an accurate estimate across a large set of states. Although this paper also contains some useful ideas and extensions for systems for which this does not hold, some parts will build on the mentioned property. In this paper, we will abuse the notation and MFPT will refer to system-wide MFPT. Values for each state will be contained in the MFPT vector.

\begin{figure}[h]
\centering
\includegraphics[width=0.48\textwidth]{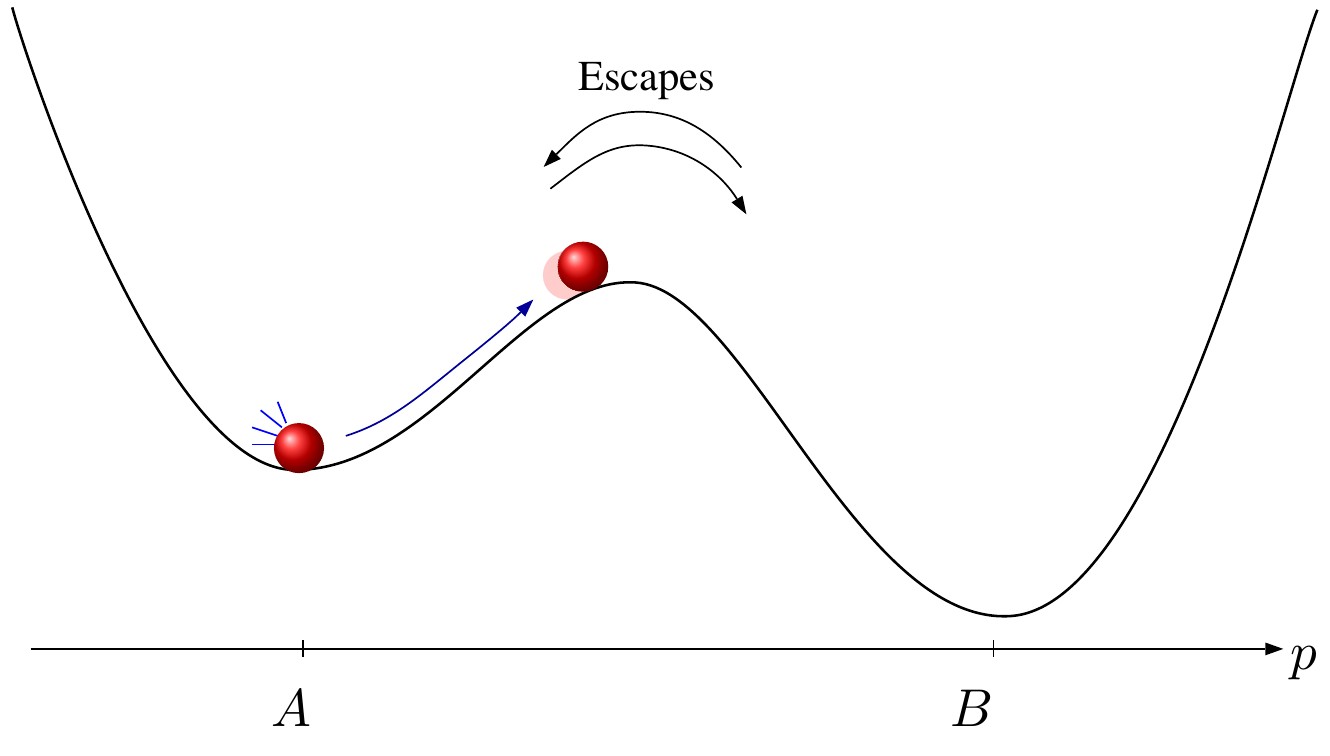}
\caption{Cartoon with multiple locally-stable equilibria under deterministic conditions. With sufficient noise in the model, the particle is guaranteed to transition from one local minimum to another (back and forth), although transitions may still be quite rare. From region $A$'s perspective, ``escape'' corresponds to moving to region $B$. Figure inspired by~\cite{byl_metastable_2009} and~\cite{benallegue_metastability_2013}.}
\label{f_ball}
\end{figure}

Of particular interest to the authors, stability of \textit{metastable} systems can potentially be well represented by (system-wide) MFPT. These systems can be natural or human-made. They exist in a precarious state of stability, appearing to be locally stable for long periods of time until an external disturbance perturbs the system into a region of state space with a qualitatively different local behavior. Since these systems are guaranteed to exit these locally well-behaved regions with probability one given enough time, they cannot be classified as ``stable'', but it is also misleading to categorize them simply as ``unstable''. A toy example of metastable systems is a ball in a hollow on terrain where a sufficient disturbance would cause the ball to roll into another local minimum, as depicted in Figure~\ref{f_ball}. Physicists have explored this phenomenon in detail and have developed a number of tools for quantifying this behavior~\cite{talkner_discrete_1987,hanggi_reaction-rate_1990,muller_rates_1997,kampen_stochastic_2011}. Metastable processes have been observed in many other branches of science and engineering including familiar systems such as crystalline structures \cite{larsen_like-charge_1997}, flip-flops \cite{veendrick_behaviour_1980}, and neuroscience \cite{fingelkurts_making_2004}.

More recently, the tools for quantifying metastable systems has been applied to walking robots to predict how a robot will perform over variable terrain for a given control policy \cite{byl_metastable_2009,saglam_robust_2014}. For such analyses, the walking robot, the environment, the system noise, and the control actions can be modeled together as a Markov chain. Assuming that the initial state of the robot lies within a so-called ``metastable region'' of state space, the eigenvalues of the state transition matrix of the Markov chain, specifically the largest eigenvalue not associated with the (absorbing) failed system state, can be used to predict the number of steps the robot can take before failing. While this approach is described in some detail in \cite{byl_metastable_2009}, we have found when building upon the basic concepts that many important aspects of the analytic approach are not immediately obvious nor (as yet) well-documented. In this paper, we attempt to clarify the overall approach and the utility of the results, using simple toy examples.

To show the applicability of our methods to various problems we consider hybrid systems, which may exhibit either or both continuous and discontinuous dynamics. The requirement is the ability to model the full dynamics as a Markov chain. This can be done in many applications where a meaningful ``step'' definition can be proposed. As we justify in subsection~\ref{sec_why_absorbing}, we will model the escape of interest (e.g., moving from region $A$ to $B$ in Fig.~\ref{f_ball}, or falling of a walking robot) as an absorbing \textit{halt} state.

In previous work, we have used the Mean First Passage Time (MFPT) to characterize the average number of Markov chain steps until reaching an absorbing failure state. In this paper, we present a more generalized concept~-- First Passage Value (FPV)~-- and discuss both the mean and \textit{variability} of a value of interest for a metastable system. Calculating different values corresponds to adopting different rewards. This was suggested in~\cite{saglam_robust_2014} and applied in~\cite{saglam_quantifying_2014}.

The rest of this paper is organized as follows. In Section~\ref{sec_motivate}, we provide some motivating examples, which we will use to illustrate for the rest of the paper. In Section~\ref{sec_absorbing_markov_chain}, we use the eigenvalues of the state transition matrix to calculate MFPT and discuss the relevance of employing a system-wide MFPT. Sections~\ref{sec_mfpv} and \ref{sec_fpv} introduces the FPV metric, which builds on MFPT. Finally, we conclude with control applications in Sec.~\ref{sec_control}.

\section{Some Motivating Examples}
\label{sec_motivate}
In this section we provide five motivating examples. The first three of them are discrete-time with no control action and can be easily represented in the form of a Markov chain. We will later discuss how to deal with more general systems in Section~\ref{sec_control}. In all these examples, we are initially interested in discrete time steps before hitting a state of interest. We will later investigate how to calculate metrics other than discrete time.

\subsection{Coin Toss}
\begin{figure}[thpb]
\centering
\begin{tikzpicture}[
align=center,node distance=.5cm,
safe/.style={circle, draw=c7!50, thick, top color =white , bottom color = c7!50,inner sep=2pt},
dangerous/.style={circle, draw=c1!50, thick, top color =white , bottom color = c1!50,inner sep=2pt},
absorbed/.style={circle, draw=c2!50, thick, top color =white , bottom color = c2!50,inner sep=2pt},
]
\def\x{2cm}
%Nodes
\node[absorbed]			(1a)														{1};
\node[dangerous]		(3a) [left=\x of 1a]										{3};
\node[safe]				(2a) [left=\x of 3a]										{2};
 
%Lines

\path (1a) edge[thick, out=+45, in=-45, loop] node[right] {0.01} (1a);
\path[->] (1a)  edge  [thick, bend left=60] node[below] {0.99} (2a);
\path (2a) edge[thick, out=180+45, in=180-45, loop] node[left] {0.99} (2a);
\path[->] (3a)  edge  [thick, bend left=20] node[below] {0.99} (2a);

\path[->] (2a)  edge  [thick, bend left=20] node[above] {0.01} (3a);
\path[->] (3a)  edge  [thick] node[above] {0.01} (1a);

\end{tikzpicture}
\caption{Markov Chain representing unfair coin toss to get two heads in a row.}
\label{f_coin}
\end{figure}
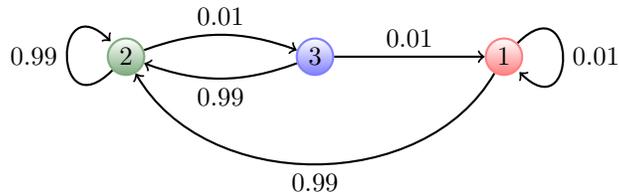

Consider tossing an unfair coin, for which the probability of having heads is $0.01$. Say we are interested in the number of flips before two heads in a row. Then, we have three possible states: (1) Heads-heads, (2) Tails in the last flip (including `not-flipped yet'), (3) Tails-Heads (including `flipped once to get heads'). Figure~\ref{f_coin} shows the corresponding Markov Chain.

\subsection{Epidemics}
\label{sec_epidemics}

Consider the discrete susceptible-infected-susceptible (SIS) model as in~\cite{ahn_global_2013}. We are interested in number of discrete time steps to everyone being healthy. As a toy example, we will consider a network of only 2 people. Then, there are four states possible as listed in Table~\ref{t_epidemics}.

{\renewcommand\arraystretch{1.5}% (MyValue=1.0 is for standard spacing)
\begin{table}[thpb]
\caption{Explanation of each state}
\label{t_epidemics}
\begin{center}
\begin{tabular}{c|c c c c}
\multicolumn{1}{c|}{}& State 1 & State 2 & State 3 & State 4\\
\hline\rowcolor{green!10}
Patient 1 & Susceptible & Susceptible & Infected & Infected \\
\rowcolor{blue!10}
Patient 2 & Susceptible & Infected & Susceptible & Infected\\
\end{tabular}
\end{center}
\end{table}}

Let `the probability of recovery when a node is infected' be $\delta=0.01$, and `the probability to be infected when the other node is infected' be $\beta=0.8$.

\subsection{Europe Tour}
\label{sec_europe}

\begin{figure}[h]
\centering
\includegraphics[width=0.48\textwidth]{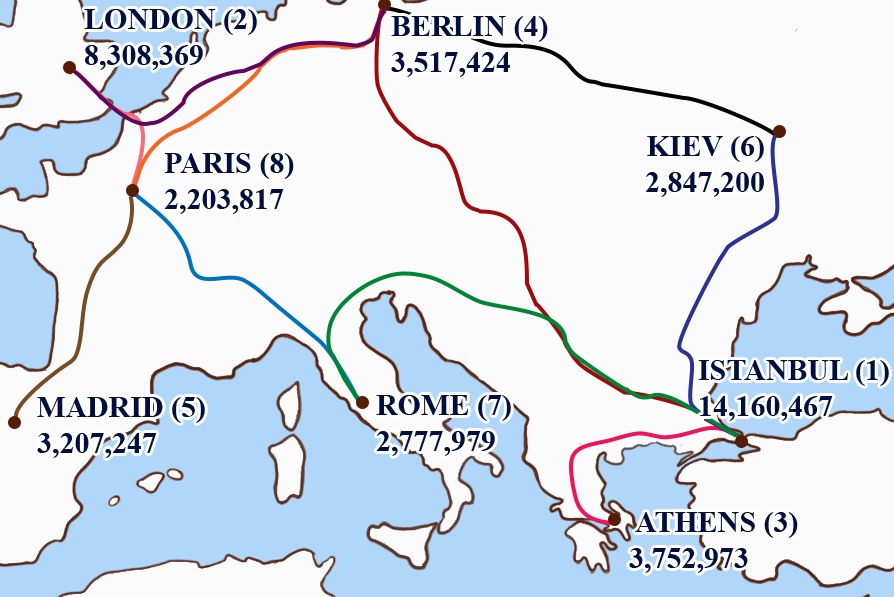}
\caption{Populations of Europe's most populated 8 cities excluding Russia. Table~\ref{t_europe} provides information about the road map drawn in this figure. Cities are ordered by their population as indicated in parenthesis, e.g., Athens (State 3) is the third most crowded city in this map.}
\label{f_roadmap}
\end{figure}

Consider a person traveling between some of the largest cities in Europe shown in Figure~\ref{f_roadmap}. After spending a day that person either stays in the same city, or moves to one of the connected cities. The probability of action is directly proportional to the population of the next city. For example when in London, this person stays there with probability 0.5922, moves to Berlin with probability 0.2507, or moves to Paris with probability 0.1571. Staying in London has the highest probability because it is more populated than Paris and Berlin. In this example, we will investigate number of days before reaching a specific city, say Istanbul.

\subsection{Legged Locomotion}
Arguably just like humans, underactuated two-legged robots are destined to fall down when the terrain is rough enough. We would like to estimate the number of steps before falling. The more steps taken on average, the more stable the robot is considered to be. Please see~\cite{saglam_robust_2015} for an application of the tools presented in this paper, where stability is dramatically increased by optimizing for number of steps.

Also, \cite{saglam_lyapunov-based_2014} provides a comparison between Lyapunov-based approach versus using the tools of this paper on a hybrid system, namely, the Rimless Wheel.

\subsection{Driving a Car}
Again arguably, given enough time (millions of years if necessary), any driver will be involved in an accident. The same is and will be true for autonomous cars including the Google car. We might be interested in, for example, the number of intersections before an accident, or the average number of miles driven between accidents. This example is provided just to motivate following sections, but we have not yet modeled it.

\section{Absorbing Markov Chains}
\label{sec_absorbing_markov_chain}
We will refer to a ``state of interest'' as a ``halt state'', e.g., two heads in a row, everybody being healthy, being in Istanbul, falling of a robot, or accident for a car. Without loss of generality, we define this halt state to be \textit{State~1} ($x_1$). Any halt state will be absorbing.

\subsection{Why Absorbing?}
\label{sec_why_absorbing}

Note that the epidemics model in the previous section is already absorbing because when both patients are susceptible (State 1), the state will not change (i.e., noone can become infected).

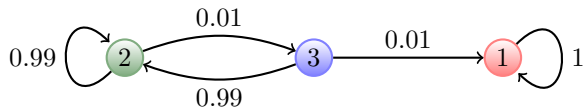
\begin{figure}[thpb]
\centering
\begin{tikzpicture}[
align=center,node distance=.5cm,
safe/.style={circle, draw=c7!50, thick, top color =white , bottom color = c7!50,inner sep=2pt},
dangerous/.style={circle, draw=c1!50, thick, top color =white , bottom color = c1!50,inner sep=2pt},
absorbed/.style={circle, draw=c2!50, thick, top color =white , bottom color = c2!50,inner sep=2pt},
]
\def\x{2cm}
%Nodes
\node[absorbed]			(1a)														{1};
\node[dangerous]		(3a) [left=\x of 1a]										{3};
\node[safe]				(2a) [left=\x of 3a]										{2};
 
%Lines

\path (1a) edge[thick, out=+45, in=-45, loop] node[right] {1} (1a);
\path (2a) edge[thick, out=180+45, in=180-45, loop] node[left] {0.99} (2a);
\path[->] (3a)  edge  [thick, bend left=20] node[below] {0.99} (2a);

\path[->] (2a)  edge  [thick, bend left=20] node[above] {0.01} (3a);
\path[->] (3a)  edge  [thick] node[above] {0.01} (1a);

\end{tikzpicture}
\caption{Figure~\ref{f_coin} modified by modeling State 1 to be absorbing.}
\label{f_coin_absorbing}
\end{figure}

Let's consider the coin toss example. State 1 is not absorbing, i.e., two heads in a row does not imply next flip to be heads. However, if we are interested in the number of flips before reaching State 1, then we should modify the Markov Chain by modeling State 1 to be absorbing as in Figure~\ref{f_coin_absorbing}. This modification corresponds to the game ending when two heads are in a row. The dynamics until State 1 do not change, but the modification allows us calculate the number of flips before State 1.

Similarly, for the Europe tour example, if we are interested in the number of days before going to Istanbul, we should model the state of being in Istanbul as absorbing.

For the following, we focus on the toy example shown in Figure~\ref{f_coin_absorbing}, which will be used to illustrate our analysis.

\subsection{The Analysis}

\label{s_markov}
The state distribution vector at step $n$ is denoted by $p[n]$ and defined by
\begin{equation}
p_i[n]:=Pr(x[n]=x_i).
\end{equation}
So $p_i[n]$ corresponds to the probability of being at state $x_i$ at step $n$. Since probability cannot be negative, $p[n]$ is a non-negative vector, and because the system has to be at a state at any step, $p[n]$ sums to 1. The state transition matrix, aka the Markov matrix or the stochastic matrix, has the following structure by definition:
\begin{equation}
T_{s\{ij\}}:=Pr(x[n+1]=x_j \ | \ x[n]=x_i).
\end{equation}
So, the element of $T_s$ on the $i$th row and $j$th column gives the probability of transitioning from state $x_i$ to state $x_j$. To illustrate, the Markov chain of Figure~\ref{f_coin_absorbing} is represented with
\begin{equation}
T_s=\begin{bmatrix}
1 & 0 & 0 
\\ 0 & 0.99 & 0.01
\\ 0.01 & 0.99 & 0
\end{bmatrix}.
\end{equation}
Similar to non-negativity of $p[n]$, we have $T_{s\{ij\}}\geq 0$. And because any state will transition (possibly to the halt state or the starting state itself) after each step, each row sums to one. For the rest of the paper, we assume the number of states is $\ell>1$. So, the state transition matrix is $\ell$ by $\ell$. The state transition matrix gives the next state distribution, given the current one:
\begin{equation}
\label{e_step_prob}
p[n+1]=T_s' \ p[n] =(T_s')^{n+1} p[0],
\end{equation}
where the prime ($'$) symbol denotes the transpose operation.

Let $\lambda$ be an eigenvalue of $T_s$. Then, there exists a non-zero vector $v$ such that
\begin{equation}
\label{e_eigen1}
T_s v=\lambda v.
\end{equation}
As shown in~\cite{matthews_number_1995}, we next show that every eigenvalue $\lambda$ of $T_s$ satisfies $|\lambda|\leq 1$. Let $k$ be such that $|v_j|\leq|v_k|$ for all $1\leq j \leq \ell$.
Equating the $k$-th components in equation~(\ref{e_eigen1}) gives
\begin{equation}
\sum_j T_{s\{kj\}} v_j=\lambda v_k.
\end{equation}
We then have
\begin{equation}
\begin{aligned}
|\lambda v_k|&=|\lambda| \ |v_k|=\bigg| \sum_j T_{s\{kj\}} v_j \bigg|
\\ & \leq \sum_j T_{s\{kj\}} |v_j| \leq \sum_j T_{s\{kj\}} |v_k|=|v_k|,
\end{aligned}
\end{equation}
where we used $T_{s\{kj\}} \geq 0$ and $\sum_j T_{s\{kj\}}=1$. $|\lambda| |v_k|\leq |v_k|$ implies $|\lambda| \leq 1$.

For the rest of the paper, we will prefer working with the transpose of $T_s$ to make the following sections easier to follow. Since $T_s$ is square, $T'_s$ has the same eigenvalues as $T_s$. Due to the nature of transpose operation and the structure of $T_s$, each column of $T'_s$ sums to one.

Remember that $x_1$ is an absorbing state, which represents the end of game, no matter how the system escaped (e.g., the robot slipped or hit the wall). Then, $T'_s$ has the following form:
\begin{equation}
\label{e_t_tilde}
T'_s=\begin{bmatrix}
1_{1\times 1} & T_1 \\
0 
& \hat{T}
\end{bmatrix}_{\ell\times \ell}.
\end{equation}
Note that $\lambda=1$ and $v=[1 \ 0 \  ... \ 0]'$ satisfies the equation
\begin{equation}
\label{e_eigen}
T'_s v=\lambda v.
\end{equation}
To distinguish (possibly non-distinct) eigenvalues, we will note them by $\lambda_j$, where $1\leq j \leq \ell$. Without loss of generality, we will let $\lambda_1=1$ and the associated basis vector be $v_1=[1 \ 0 \  ... \ 0]'$.

Existence of Jordan normal form for any square matrix is fundamental to Linear Algebra. Consider a Jordan normal form of $\hat{T}$ given by
\begin{equation}
\label{e_jordan}
\hat{T}=\hat{V}\hat{J}\hat{V}^{-1},
\end{equation}
Then, as we will verify, a Jordan normal form of $T'_s$ is given by
\begin{equation}
\label{e_big_jordan}
T'_s=VJV^{-1},
\end{equation}
\begin{equation}
\label{e_big_jordan_j}
\text{where }J=\begin{bmatrix}
1 & 0
\\ 0 & \hat{J}
\end{bmatrix},
\end{equation}
\begin{equation}
\label{e_v_tilde}
\text{and }V=\begin{bmatrix}
1 & -[1\ ... \ 1]\hat{V}
\\ 0 & \hat{V}
\end{bmatrix}.
\end{equation}
Note that the sum of each column of $V$ equals zero, except the first one. Furthermore, these columns form a basis in $\mathbb{R}^\ell$. Equation~(\ref{e_big_jordan}) can be verified as follows. The inverse of $V$ is
\begin{equation}
V^{-1}=\begin{bmatrix}
1 & [1\ ... \ 1]
\\ 0 & \hat{V}^{-1}
\end{bmatrix}.
\end{equation}
Then, the right hand side of (\ref{e_big_jordan}) can be calculated as
\begin{equation}
\begin{bmatrix}
1 & [1\ ... \ 1]-[1\ ... \ 1]\hat{V}\hat{J}\hat{V}^{-1}
\\ 0 & \hat{V}\hat{J}\hat{V}^{-1}
\end{bmatrix}=\begin{bmatrix}
1 & [1\ ... \ 1](I-\hat{T})
\\ 0 & \hat{T}
\end{bmatrix}.
\end{equation}
Equation~(\ref{e_big_jordan}) is thus verified because $T_1+[1\ ... \ 1]\hat{T}=[1\ ... \ 1]$ (columns of $T'_s$ sum to one).

The spectrum of $\hat{T}$, denoted by $\sigma(\hat{T})$, is the set of distinct eigenvalues of $\hat{T}$. The spectral radius of $\hat{T}$ is given by
\begin{equation}
\rho(\hat{T})=\max\limits_{\lambda \in \sigma(\hat{T})} |\lambda|.
\end{equation}
Let the spectral radius be $r=\rho(\hat{T})$. Remembering that $\hat{T}$ is a non-negative square matrix, we present the following two facts, which are proven in~\cite{meyer_matrix_2000}.
\begin{enumerate}
  \item $r\in \sigma(\hat{T})$ (i.e., $r$ is an eigenvalue of $\hat{T}$).
  \item $\hat{T}z=rz$ for some $z\in \mathcal{N}=\{v| \ v\geq 0 \text{ with } v\neq 0 \}$
\end{enumerate}
We will let $\lambda_2:=r$ and $v_2$ will refer to the associated column in $V$. Then, $v_2=\left[-\norm{z}_1 \ z'\right]'$. Now, let us consider the following state distribution
\begin{equation}
\phi:=\left[ 0 \ \ \frac{z'}{\norm{z}_1} \right]'=v_1+\frac{1}{\norm{z}_1}v_2.
\end{equation}
We will call $\phi$ the \textit{metastable distribution}. Note that it is a valid initial state distribution since it sums to one and each element is non-negative. Back to our toy example, we have
\begin{equation}
\label{e_toy_phi}
\lambda_2\approx 1-9.9020\times 10^{-5} \text{ and } \phi\approx\begin{bmatrix}
0
\\0.9901
\\0.0099
\end{bmatrix}
\end{equation} 
This metastable distribution represents a simple probability distribution: The state is $x_2$ with probability $\approx$0.9901 and it is $x_3$ with probability $\approx$0.0099. The first element of $\phi$, i.e., the probability of being at $x_1$, is zero, by definition (of not yet having escaped). It is actually interesting that the second element in $\phi$ is higher than the probability of having tails in a flip. Yet, this is true as can be verified by exactly solving for $\phi$ and $\lambda_2$ using the method we will show shortly.

Taking a step when $\phi$ is the initial condition, we obtain
\begin{equation}
T'_s \phi =T'_s \left(v_1+\frac{1}{\norm{z}_1}v_2\right)=v_1 +\frac{\lambda_2}{\norm{z}_1}v_2.
\end{equation}
Naturally, the resulting distribution is also non-negative and sums to one. In addition, the first element is
\begin{equation}
1+\frac{\lambda_2}{\norm{z}_1}(-\norm{z}_1)=1-\lambda_2.
\end{equation}
This means the system escaped with probability $1-\lambda_2$. Furthermore, given the system did not escape, we also have $\phi$ as the final probability distribution for the states. This can be seen by zeroing the first element of $T'_s \phi$ and scaling to sum to one. Then we have the following result: When $\phi$ is the initial state distribution, the probability of escaping is $1-\lambda_2$, the probability of staying in the same distribution ($\phi$) is $\lambda_2$. In our toy example, these can be seen by calculating

\begin{equation}
\label{e_toy_solve}
T_s'\phi\approx \begin{bmatrix}
0.0001
    \\0.9900
    \\0.0099
\end{bmatrix} \text{ and } \frac{T_s'\phi-(1-\lambda_2)
\begin{bmatrix}
1
    \\0
    \\0
\end{bmatrix}}{\norm{ T_s'\phi-(1-\lambda_2)
\begin{bmatrix}
1
    \\0
    \\0
\end{bmatrix} }_1}=\phi.
\end{equation}
Note that $\lambda_2\approx 1-9.9020\times 10^{-5}$ is close to one. This will be the case (actually $1-\lambda_2$ will be much smaller in many cases) when we are interested in absorbing Markov chains with ``rare escapes''.

For this toy example, we can write $\phi$ and $\lambda_2$ in exact form as mentioned. This is done simply by solving the right hand side of~(\ref{e_toy_solve}) and noting $\norm{\phi}_1=1$. In practice, a numerical eigenvalue solution gives fast and accurate results, even for quite complex (e.g., 411,000 states in~\cite{byl_metastable_2009}) models.

We then calculate the average number of steps before escape, given the initial condition was $\phi$, i.e. $p[0]=\phi$. This corresponds to the term Mean First Passage Time (MFPT) in~\cite{byl_metastable_2009}. The higher MFPT is, the more stable a system is said to be. There are two cases depending on the probability of taking a step: If $\lambda_2=1$, then the probability of escape is zero. In this case the system will take infinitely many steps without escaping to the halt state. The other case ($\lambda_2<1$) is relatively more complicated and interesting as explained next. Note that from this point on we'll focus on the case $\lambda_2<1$.

\subsection{(System-wide) First Passage Time (FPT)}

Given the probability of taking a step without escaping is $\lambda_2<1$, the probability of taking $n$ steps only, equivalently escaping at the $n$th step is simply
\begin{equation}
Pr(x[n]=x_1, \ x[n-1]\neq x_1)=\lambda_2^{n-1}(1-\lambda_2).
\end{equation}
Realize that as $n\to\infty$, the right hand side goes to zero, i.e., the system will eventually escape. Note that we also count the step which ended up escaping as a step. This can be verified considering escaping at the first step (taking 1 step only). When $n=1$ is substituted, we get $1-\lambda_2$ as expected. Then, the expected (mean) number of steps can be calculated as 

\begin{equation}
\label{e_mfpt}
\begin{aligned}
MFPT & = E[FPT] 
\\ & = \sum\limits_{n=1}^{\infty} n \ Pr(x[n]=x_1, \ x[n-1]\neq x_1)  
\\ &  = \sum\limits_{n=1}^{\infty}n \lambda_2^{n-1} (1-\lambda_2)
= \frac{1}{1-\lambda_2},
\end{aligned}
\end{equation}
where we used the fact that $\lambda_2<1$. As a result, MFPT can then be calculated using
\begin{equation}
M=\begin{cases}
\infty & \lambda_2=1, \\
\frac{1}{1-\lambda_2} & \lambda_2<1.
\end{cases}
\end{equation}

The MFPT of our toy example is
\begin{equation}
\frac{1}{1-\lambda_2}\approx 1.0099\times 10^4.
\end{equation}
So, in the toy example if we start with the initial state distribution $\phi$ given in (\ref{e_toy_phi}), then the system will take approximately 10,099 steps on average.

The standard deviation of FPT can be calculated by
\begin{equation*}
\begin{aligned}
E[FPT^2] &= \sum\limits_{n=1}^{\infty} n^2 \ Pr(x[n]=x_1, \ x[n-1]\neq x_1)
\\ &  = \sum\limits_{n=1}^{\infty}n^2 \lambda_2^{n-1} (1-\lambda_2) =
\frac{1+\lambda_2}{(1-\lambda_2)^2}
\end{aligned}
\end{equation*}
\vspace{.2cm}
\begin{equation}
\implies
\sqrt{E[FPT^2]-(E[FPT])^2}=M\sqrt{\lambda_2}.
\end{equation}
This corresponds to $M\sqrt{\lambda_2}\approx 10,098.5$. For our toy example, as for any metastable system, $\lambda_2$ being close to one results in a standard deviation close to the mean!

We are also interested in obtaining the MFPT vector, $m$, which gives the MFPT for each state.
\begin{equation}
\label{e_mfpt_vector1}
m_i:=\begin{cases}
0 & i=1,\\
1+\sum\limits_{j} T_{s\{ij\}}m_j & otherwise.
\end{cases}
\end{equation}
The equation above says it will take zero steps to go to the halt state if the system escaped already. Otherwise, the number of steps until halt is 1 less after a step is taken. For the toy example, this means $m_1=0$, $m_2=1+0 m_1+0.99 m_2+0.01 m_3$, and $m_3=1+0.01 m_1+0.99 m_2+0 m_3$. These equations can be solved to get
\begin{equation}
m=\begin{bmatrix}
0
\\1.01\times10^4\\10^4
\end{bmatrix}.
\end{equation}
This means if we start at state $x_2$, we expect to take $1.01\times10^4$ steps before escape.

By using~(\ref{e_t_tilde}), it is straightforward to obtain
\begin{equation}
\label{e_mfpt_vector}
m=\begin{bmatrix}
0\\
(I-\hat{T}')^{-1}1
\end{bmatrix}.
\end{equation}
To be able to calculate (\ref{e_mfpt_vector}), we need $(I-\hat{T}')$ to be invertible. This is equivalent to having $\lambda_2<1$, which is the hidden assumption we made while defining the MFPT vector. The system-wide MFPT calculated in~(\ref{e_mfpt}) can be also obtained by
\begin{equation}
M=m'\phi=\frac{1}{\norm{z}_1}[1\ ... \ 1] (I-\hat{T})^{-1}z.
\end{equation}
This makes sense because each state has its own MFPT, and MFPT of the metastable distribution is just a convex combination of each state's MFPT weighted according to $\phi$. We show the equivalence next.
\begin{equation}
\begin{aligned}
\hat{M} & =\frac{1}{\norm{z}_1}[1\ ... \ 1] (I-\hat{T})^{-1}z
\\ & =\frac{1}{\norm{z}_1}[1\ ... \ 1] (I-\hat{T})^{-1}(I-\hat{T}+\hat{T})z
\\ & =\frac{1}{\norm{z}_1}[1\ ... \ 1] (I+(I-\hat{T})^{-1}\hat{T})z
\\ & =\frac{1}{\norm{z}_1}[1\ ... \ 1]z+ \frac{1}{\norm{z}_1}[1\ ... \ 1](I-\hat{T})^{-1}\hat{T}z
\\ & =1+ \lambda_2 \frac{1}{\norm{z}_1}[1\ ... \ 1](I-\hat{T})^{-1}z
\\ & =1+ \lambda_2 \hat{M} \ \ \implies \hat{M}=1/(1-\lambda_2)=M
\end{aligned}
\end{equation}
Note that $M$ is upper bounded by the largest element in $m$. In fact, any initial state distribution, $p[0]$, will have an MFPT that is a convex combination of the $m_i$ values.

\subsection{Why (system-wide) MFPT?}
\label{sec_why}

We would like to answer why we should be using MFPT of the metastable distribution. First of all, it is a lower bound for average steps taken from at least one of the states, because it is a convex combination of $m_i$s. So, there are state(s) at least as stable. Secondly, it is a good measure of overall stability. Often systems quickly converge to their metastable distributions, where MFPT becomes the true value. Thirdly, system-wide MFPT also has advantages over calculating the MFPT vector. In case $T_s$ is very large, estimating the second largest eigenvalue is relatively very easy, whereas finding the inverse to calculate MFPT vector costs more time. Also, a scalar representing the stability is much easier to understand than a possibly huge vector.

Let's assume $(T_s)$ has distinct eigenvalues. Please see~\cite{saglam_metastable_2014} for other cases. Let the initial distribution be
\begin{equation}
p[0]=c_1v_1+c_2v_2+...c_{\ell}v_{\ell}.
\end{equation}
Note that $c_1=1$ to have $\norm{p[0]}_1=1$. Then,
\begin{equation}
p[n]=(T_s')^np[0]=v_1+c_2\lambda_2^nv_2+...+c_{\ell}\lambda_{\ell}^nv_{\ell}.
\end{equation}

In the light of this section we see that the metastable distribution is also given by
\begin{equation}
\label{e_phi}
\phi_i=\lim\limits_{n\to\infty}Pr(X[n]=x_i \ | X[n]\neq x_1]),
\end{equation}
when the limit exists. (It does for distinct eigenvalues)

\subsection{How quickly is the Initial Distribution Forgotten?}
\label{sec_memory}
First, let's explain what we mean by being ``forgotten''. We say the initial distribution (condition) is forgotten if either the distribution is the metastable distribution ($p=\phi$) or the game ended (halt state) ($p=[1 \ 0 \  ... \ 0]'$). So, the question can be paraphrased as ``how quickly do we converge towards the metastable distribution, given the system is not absorbed yet?''. For systems with distinct eigenvalues, we propose using

\begin{equation}
\text{memory constant}=\frac{1-\lambda_2}{1-|\lambda_3|}.
\end{equation}

We are motivated by the fact that $\sum_{n=0}^{\infty}\lambda_i^n=1/(1-\lambda_i)$ for $|\lambda_i|<1$. Although this holds for complex $\lambda_i$ values too, since $|\lambda_i^n|=|\lambda_i|^n$, we use $1/(1-|\lambda_i|)$ to quantify how many steps it takes before vanishing. The higher $\lambda_3$ is, the slower the initial condition is forgotten, i.e., the more steps are required to forget the same amount of initial condition information ($1/(1-|\lambda_3|)$ is higher). We look at $\lambda_3$, because it gives the worst case scenario, i.e., it gives a conservative value. We divide ($1/(1-|\lambda_3|)$ by $M=1/(1-\lambda_2)$ to get a relative memory constant, which is upper bounded by 1.

It is also worth noting that to get a memory constant of $10^{-6}$, we need $\lambda_2=1-10^{-6}(1-|\lambda_3|)>1-10^{-6}$, or equivalently MFPT $M>10^6$. Thus, small memory constants require metastability.

If the memory constant is very close to one, then we have another mode almost as stable (as the one associated with $\lambda_2$). In such cases it may be useful to use the next $|\lambda_i|$ instead of $|\lambda_3|$, until we have a memory much smaller than one.

When the eigenvalues are not necessarily distinct, we may have $\lambda_2=|\lambda_3|$. This case is very similar to the one just explained.

\subsection{Epidemics}
For the epidemics model outlined in Section~\ref{sec_epidemics}, the stochastic matrix is given by

\begin{equation}
\begin{aligned}
T_s&=\begin{bsmallmatrix}
1 & 0 & 0 & 0
\\ (1-\beta)\delta & (1-\delta)(1-\beta) & \beta\delta & \beta(1-\delta)
\\ (1-\beta)\delta & \beta\delta & (1-\delta)(1-\beta) & \beta(1-\delta)
\\ \delta\delta & \delta(1-\delta) & \delta(1-\delta) & (1-\delta)(1-\delta)
\end{bsmallmatrix}\\
&=\begin{bmatrix}
1 &        0    &     0     &    0\\
    0.002  &  0.198   & 0.008 &   0.792\\
    0.002  &  0.008  &  0.198  &  0.792\\
    0.001  &  0.0099 &   0.0099   & 0.9801
\end{bmatrix}.
\end{aligned}
\end{equation}
$\lambda_2$ is such that MFPT is $M=6.8383\times10^3$. In addition, $\lambda_3=0.19$, and the memory constant is $1.8\times10^{-4}$. Small memory constant results in states having a MFPT close to either zero or $M$.

\begin{equation}
m=\begin{bmatrix}
0\\
6822.7\\
    6822.7\\
    6838.7
\end{bmatrix}
\hspace{1cm}
\phi=\begin{bmatrix}
0\\
    0.0122\\
    0.0122\\
    0.9757
\end{bmatrix}
\end{equation}

\subsection{Europe Tour}
We now look at the example provided in Section~\ref{sec_europe}. We have $\lambda_2=0.866$ and $\lambda_3=0.6355$. As a result MFPT is $M=7.4643$ and the memory constant is 0.3676. Due to the high memory constant, we see a high variation between the MFPT of each state.

\begin{equation}
m=\begin{bmatrix}
0\\
8.4084\\
    1.265\\
    4.5381\\
   10.6749\\
    2.064\\
    2.2767\\
    8.2196
\end{bmatrix}
\hspace{1cm}
\phi=\begin{bmatrix}
0\\
0.4620\\
    0\\
    0.2006\\
    0.1028\\
    0.0253\\
    0.0338\\
    0.1754
\end{bmatrix}
\end{equation}

As an interesting fact, note that there is a zero probability of being in Athens in the metastable distribution, because you can only go to Athens from Istanbul, which is the absorbing state.

\subsection{Modified Europe Tour}
\label{sec_mod_europe}
Since metastable systems with a small memory constant are of interest to the authors, we modify the Europe tour example by hypothetically assuming the population of Paris is 1 billion instead. Then, we have a MFPT of $M=10,823$ and the memory constant drops to $1.1688\times10^{-4}$. This results in system-wide MFPT being more ``valid'', because most of the states either have a MFPT close to either zero or $M$.

\begin{equation}
m=\begin{bmatrix}
0\\
10,825\\
    1\\
    10,652\\
    10,825\\
    2,121\\
    10,674\\
    10,824\\
\end{bmatrix}
\hspace{1cm}
\phi=\begin{bmatrix}
0\\
0.0081\\
   0\\
    0.0034\\
    0.0031\\
    0\\
    0.0027\\
    0.9826
\end{bmatrix}
\end{equation}
This time in addition to Athens, Kiev also no longer appears in the metastable distribution. This is because within several steps we will typically move directly from Kiev to either Istanbul (the absorbing state) or Berlin. The latter almost implies going to Paris in the following step due to high population there. Kiev has a MFPT that is not very small or close to the system-wide MFPT, because $m_1=0$, $m_4\approx M$, and
\begin{equation}
m_6=T_{s\{61\}}m_1+T_{s\{64\}}m_4+T_{s\{66\}}m_6
\end{equation}
implies
\begin{equation}
m_6\approx\frac{T_{s\{64\}}}{1-T_{s\{66\}}}=2,153.
\end{equation}

\section{Measuring Metrics Other Than Steps}
\label{sec_mfpv}
Calculating the discrete time to a state of interest has potential to be very useful, but in some applications we might be interested in metrics other than discrete time. To illustrate, we will build on the modified Europe Tour of Section~\ref{sec_mod_europe}. Instead of how many decisions it took before reaching Istanbul, we might be interested in travel-time or distance to Istanbul. For this matter, we propose the term Mean First Passage Value (MFPV), where ``value'' depends on the task, e.g., Mean First Passage Distance.

\subsection{Mean First Passage Value (MFPV)}
First, note that Mean First Passage Value (MFPV) is a generalization of MFPT. Equivalently, MFPT is a special case of MFPV, where value is discrete time steps.

Let us start by redefining $m$ from (\ref{e_mfpt_vector1}) as the MFPV vector, which gives the MFPV for each state, as
\begin{equation}
\label{e_mfpv_vector1}
m_i:=\begin{cases}
0 & i=1\\
\sum\limits_{j} T_{s\{ij\}} T_{v\{ij\}}+\sum\limits_{j} T_{s\{ij\}}m_j & otherwise,
\end{cases}
\end{equation}
where $T_{v\{ij\}}$ is the value (reward) of transitioning from state $x_i$ to $x_j$. For example, travel from Rome to Paris takes 12 hours 46 minutes. Then, we can calculate vector $m$ as
\begin{equation}
\label{e_mfpv_vector}
m=\begin{bmatrix}
0\\
(I-\hat{T}')^{-1}
\begin{bmatrix}
\vspace{-.25cm}\\
\sum\limits_{j} T_{s\{2j\}} T_{v\{2j\}}\\
.\\
.\\
.\\
\sum\limits_{j} T_{s\{\ell j\}} T_{v\{\ell j\}}
\end{bmatrix}
\end{bmatrix}.
\end{equation}
And the (system-wide) MFPV is simply
\begin{equation}
M=m'\phi.
\end{equation}

\subsection{Modified Europe Tour}
For the modified Europe tour example and the information provided in Table~\ref{t_europe}, Mean First Passage Distance is 325.68 thousand km. This is achieved by setting, for example, $T_{v\{24\}}=1,098$ km, which corresponds to the distance between London and Berlin. 

{\renewcommand\arraystretch{1.5}% (MyValue=1.0 is for standard spacing)
\begin{table}[thpb]
\caption{Explanation of each state}
\label{t_europe}
\begin{center}
\begin{tabular}{c|c c }
\multicolumn{1}{c|}{}& Time (h:m) & Distance (km) \\
\hline\rowcolor{green!10}
London (2) - Paris (8) & 5:06 & 454  \\
\rowcolor{blue!10}
London (2) - Berlin (4) & 10:25 & 1,098  \\
\rowcolor{green!10}
Madrid (5) - Paris (8) & 11:10 & 1,270  \\
\rowcolor{blue!10}
Rome (7) - Paris (8) & 12:46 & 1,419  \\
\rowcolor{green!10}
Berlin (4) - Paris (8) & 9:18 & 1,055  \\
\rowcolor{blue!10}
Berlin (4) - Kiev (6) & 14:41 & 1,329  \\
\rowcolor{green!10}
Berlin (4) - Istanbul (1) & 21:39 & 2,210  \\
\rowcolor{blue!10}
Kiev (6) - Istanbul (1) & 19:00 & 1,459  \\
\rowcolor{green!10}
Rome (7) - Istanbul (1) & 22:46 & 2,262  \\
\rowcolor{blue!10}
Athens (3) - Istanbul (1) & 11:13 & 1,095  \\
\end{tabular}
\end{center}
\end{table}}

To calculate the travel-time only (excluding the days spent in a city), we set $T_{v\{ii\}}=0$. Then, it takes 128 days on average before we are in Istanbul. The MFPV vector for this case is
\begin{equation}
\label{e_mfpv_vector}
m=\begin{bmatrix}
0\\
128.4756\\
    0.4674\\
  126.6098\\
  128.7334\\
   25.9478\\
  127.0149\\
  128.2681
\end{bmatrix} \text{ days.}
\end{equation}

On the other hand, if we also include the days spent in a city (remember that a decision is made after spending a day in the city), we need to set $T_{v\{ii\}}=1$ day to obtain MFPV of 29 years, which is much higher than 128 days. One reason of this is because once in Paris, there is a 0.9825 probability to stay in Paris.

\subsection{The Value}
Often, one wishes to include multiple objectives in a single value function, for example for walking robots penalizing failure events (e.g., falling down) while also rewarding fast speed and low energy use. This can be achieved since the value (i.e., cost function) does not need to have a physical correspondence, number of steps minus $10^{-3}$ times energy consumption is a valid value definition. Please see~\cite{saglam_quantifying_2014} for further details and usage of this example.

\section{Confidence Levels on Value}
\label{sec_fpv}
In some applications, instead of the mean FPV, we may want to have a conservative FPV bound for a particular ``confidence level'', $pr$. That is, one would observe a value above LFPT with probability $pr$, and one would observe a value below UFPT with probability $pr$.

\subsection{First Passage Time (FPT)}
The probability of taking more than LFPT steps is $\lambda_2^{\text{LFPT}}$. Then, the lower bound on number of steps taken with probability $pr$ can be calculated by
\begin{equation}
\text{LFPT}(pr)=log_{\lambda_2} pr.
\end{equation}
Note that LFPT(1)=0, so we can only guarantee taking a single step, which leads to the halt state. The probability of taking less than UFPT steps is $1-\lambda_2^{\text{UFPT}-1}$. Then, the upper bound on number of steps taken with probability $pr$ is
\begin{equation}
\text{UFPT}(pr)=log_{\lambda_2} (1-pr)+1.
\end{equation}
Note that $\lim\limits_{pr\to 1}\text{UFPT}(pr)=\infty$, so it may take infinitely many steps before converging to the halt state (theoretically). We can then limit the FPT for a given probability
\begin{equation}
\text{LFPT}\leq \text{FPT} \leq \text{UFPT}.
\end{equation}
To have LFPT$<$UFPT we require $pr>\lambda_2/(1+\lambda_2)$, which is approximately 0.5 for $\lambda_2\approx 1$. To illustrate the advantage of looking to FPT, let's consider the modified Europe tour example. The probability of the journey taking more than $M=10,823$ days is only 36.79~\%. On the other hand, we have
\begin{equation}
\begin{aligned}
1,140\leq \text{FPT}(0.9)\leq 24,921
\\108\leq \text{FPT}(0.99)\leq 49,842.
\end{aligned}
\end{equation}
These numbers are not coincidence. As $\lambda_2 \to 1$, the probability of taking $M$ steps is
\begin{equation}
\lim\limits_{\lambda_2\to 1}\lambda_2^{1/(1-\lambda_2)}=\frac{1}{e}\approx 0.3679.
\end{equation}
In fact, $0.3679$ is an upper limit for the probability of taking $1/(1-\lambda_2)$ steps for any $\lambda_2$. In addition, when $\lambda_2$ and $pr$ are close to one, we have
\begin{equation}
\text{LFPT}(pr)=\frac{ln(pr)}{ln(\lambda_2)}\approx\frac{1-pr}{1-\lambda_2}=(1-pr)M,
\end{equation}
\begin{equation}
\text{UFPT}(pr)=\frac{ln(1-pr)}{ln(\lambda_2)}+1\approx-ln(1-pr)M,
\end{equation}
where $M$ denotes MFPT.

Note that LFPT is of interest for walking systems, to give a conservative bound on steps to failure, while UFPT would be helpful in modeling epidemics, to get a conservative time when everyone will be healthy (i.e., recurrence to an ``all-healthy'' system state).

\subsection{First Passage Value}
Using MFPT and MFPV we can calculate value per step. We then, simply use
\begin{equation}
\text{FPV}=\frac{\text{MFPV}}{\text{MFPT}}\text{ FPT}.
\end{equation}
When the value is the distance in the modified Europe tour example, we obtain
\begin{equation}
\begin{aligned}
3.4 \times 10^4\leq \text{FPV}(0.9)\leq 7.5 \times 10^{5}
\\3.2\times 10^3 \leq \text{FPV}(0.99)\leq 1.5\times 10^{6}.
\end{aligned}
\end{equation}
So, with probability 0.99, the journey will take more than 3.2 thousand or less than 1.5 million kilometers.

\section{Control Applications}
\label{sec_control}
In order to control a system, one needs a goal, e.g., for the Europe tour example, trying to go to Istanbul as quickly as possible. It is then useful to quantify the performance towards achieving that goal, e.g., number of days before reaching Istanbul would be a meaningful metric to minimize. We believe FPV tool would give useful metrics for many applications. To illustrate, the more steps a walking robot takes before falling, the more stable it is said to be. Once we quantify the performance, we can optimize the control.

For the examples we studied in the previous sections, there was no control. We just illustrated how to calculate FPV for a given Markov Chain. There are two ways control action comes into play: Low-level and high-level. Let's explain by illustrating with the Europe tour. Say the probability for city change is directly proportional to population to the power $k$, where $1\leq k\leq 3$. What we previously studied was a special case of this setting with $k=1$. Any choice of $k$ will give a Markov Chain, for which we calculate FPV as shown. Choosing $k$ is the low-level control problem. Moreover, if we are capable of choosing an integer $k$ every time a city is to be chosen, then we have three low-level controllers available. Deciding which one to use for a given city is the high-level control problem.

Remember that the first three examples in Section~\ref{sec_motivate} were all discrete-time. So, they were special cases of hybrid systems, which may have either or both discontinuities and continuous phases. For systems in which an exact (discrete) Markov chain does not naturally exist, we can create one, using the meshing process described below.

\subsection{Hybrid Model}
Let $x$, $\gamma$, and $\zeta$ be the internal state, the randomness system experiences, and the control action respectively. To illustrate, for a walking robot, $x$ is the robot's state, $\gamma$ is random variable representing factors such as terrain variation or system noise, and $\zeta$ is the control action which may be a function of $x$ and $\gamma$. We define vector $y:=[x;\gamma;\zeta]$ to represent them all. Then our general hybrid model is represented as
\begin{equation}
\begin{aligned}
\dot{y}&=f(y) \hspace{.6cm} &y\in C
\\y^+&=g(y) & y\in D.
\end{aligned}
\end{equation}
$C$ and $D$ are known as flow and jump sets~\cite{goebel_hybrid_2012}. Note that this setting is compatible with less general cases like continuous and discrete systems with/without a control action or randomness.

\subsection{Meshing for Markov Decision Process (MDP) Model}
\label{sec_meshing}
The first step is choosing a Poincar\'{e}-like section, noted by $S$, which does not necessarily decrease the dimension of the state. However, if the system has not yet escaped (from the region of interest), it needs to keep passing through this section. For example, the hybrid dynamics of walking systems are punctuated by discrete impacts when a foot comes into contact with the ground. These impacts provide a natural discretization of the robot motion.

After defining this section, we abuse the notation and refer to $x\in S$ simply by $x$. Then, the next state (intersecting $S$) is a function of the current state $x[n]$, the randomness experienced $\gamma[n]$, and the controller action during that step $\zeta[n]$ i.e.,
\begin{equation}
\label{e_first}
x[n+1]=h(x[n],\gamma[n],\zeta[n]).
\end{equation}

To obtain a (discrete) Markov Decision Process (MDP) model, we need to have finite sets for control action, randomness, and state. The first one is rather easy, we simply design finitely many low-level controllers. The second (randomness), is straightforward to handle when number of noise sources is low, e.g., when randomness is in the slope ahead for a walking robot, randomness set is just one dimensional, which is very easy to discretize by meshing. State space can be discretized similarly when the state is low dimensional. However, if the state is high dimensional, discretization is not as intuitive. In case the intrinsic dimension is low, we can still mesh by cleverly exploring the reachable state space, as explained in~\cite{saglam_robust_2015}.

\begin{figure}[thpb]
\centering
\begin{tikzpicture}[
align=center,node distance=.5cm,
safe/.style={circle, draw=c7!50, thick, top color =white , bottom color = c7!50,inner sep=2pt},
dangerous/.style={circle, draw=c1!50, thick, top color =white , bottom color = c1!50,inner sep=2pt},
absorbed/.style={circle, draw=c2!50, thick, top color =white , bottom color = c2!50,inner sep=2pt},
]
\def\x{.6cm}
%Nodes
\node[absorbed]			(1a)														{1};
\node[dangerous]		(3a) [left=\x of 1a]										{3};
\node[safe]				(2a) [left=\x of 3a]										{2};
\node [above=.5cm of 3a] {\textbf{$(\zeta_1,\gamma_1)$}};

\node[absorbed]			(1b) [right=3.5cm of 1a]													{1};
\node[dangerous]		(3b) [left=\x of 1b]										{3};
\node[safe]				(2b) [left=\x of 3b]										{2};
\node [above=.5cm of 3b] {\textbf{$(\zeta_1,\gamma_2)$}};

\node[absorbed]			(1c)	[below=2cm of 1a]													{1};
\node[dangerous]		(3c) [left=\x of 1c]										{3};
\node[safe]				(2c) [left=\x of 3c]										{2};
\node [above=.5cm of 3c] {\textbf{$(\zeta_2,\gamma_1)$}};

\node[absorbed]			(1d)	[below=2cm of 1b]													{1};
\node[dangerous]		(3d) [left=\x of 1d]										{3};
\node[safe]				(2d) [left=\x of 3d]										{2};
\node [above=.5cm of 3d] {\textbf{$(\zeta_2,\gamma_2)$}};
 
%Lines
%\coordinate (Middle) at ($(1a)!0.5!(1b)$);
%\coordinate (Middle2) at ($(1c)!0.5!(1d)$);
%\draw [dashed] (Middle) -- (Middle2);
\draw [dashed] (-2,-1) -- (4,-1);
\draw [dashed] (1.2,1) -- (1.2,-3);

\path (1a) edge[thick, out=+45, in=-45, loop] (1a);
\path[->] (2a)  edge  [thick, bend left=40] (1a);
\path[->] (3a)  edge  [thick] (2a);

\path (1b) edge[thick, out=+45, in=-45, loop] (1b);
\path[->] (2b)  edge  [thick] (3b);
\path[->] (3b)  edge  [thick] (1b);

\path (1c) edge[thick, out=+45, in=-45, loop] (1c);
\path (2c) edge[thick, out=180+45, in=180-45, loop] (2c);
\path[->] (3c)  edge  [thick] (1c);

\path (1d) edge[thick, out=+45, in=-45, loop] (1d);
\path[->] (2d)  edge  [thick, bend left=40] (1d);
\path[->] (3d)  edge  [thick] (1d);
\end{tikzpicture}
\caption{Representation of a Markov Decision Process with two available actions and two possible randomness}
\label{f_1}
\end{figure}
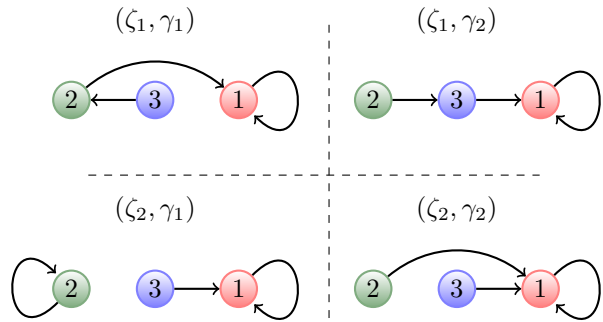

Once we have finite control, randomness and state sets, we simply simulate for each possible $h(x,\gamma,\zeta)$ to obtain a MDP similar to Figure~\ref{f_1}, where $\gamma[n]\in\{\gamma_1,\gamma_2\}$ and $\zeta[n]\in\{\zeta_1,\zeta_2\}$, i.e., there are two available actions (low-level control) and two possible randomness. Also there are just 3 states, that is $x[n]\in\{x_1,x_2,x_3\}$.

\subsection{Policy for $\zeta$}
\begin{figure}[thpb]
\centering
\begin{tikzpicture}[
align=center,node distance=.5cm,
safe/.style={circle, draw=c7!50, thick, top color =white , bottom color = c7!50,inner sep=2pt},
dangerous/.style={circle, draw=c1!50, thick, top color =white , bottom color = c1!50,inner sep=2pt},
absorbed/.style={circle, draw=c2!50, thick, top color =white , bottom color = c2!50,inner sep=2pt},
]
\def\x{.6cm}

%Lines
\coordinate (Middle) at (1.3,1);
%\draw [dashed] (Middle)+(0,2) -- (Middle);
\draw [dashed] (1.2,1) -- (1.2,-1);

%Nodes
\node[absorbed]			(1a)														{1};
\node[dangerous]		(3a) [left=\x of 1a]										{3};
\node[safe]				(2a) [left=\x of 3a]										{2};
\node [above=.5cm of 3a] {\textbf{$(\pi,\gamma_1)$}};

\node[absorbed]			(1b) [right=3.5cm of 1a]														{1};
\node[dangerous]		(3b) [left=\x of 1b]										{3};
\node[safe]				(2b) [left=\x of 3b]										{2};
\node [above=.5cm of 3b] {\textbf{$(\pi,\gamma_2)$}};

\path (1a) edge[thick, out=+45, in=-45, loop] (1a);
\path (2a) edge[thick, out=180+45, in=180-45, loop] (2a);
\path[->] (3a)  edge  [thick] (2a);

\path (1b) edge[thick, out=+45, in=-45, loop] (1b);
\path[->] (2b)  edge  [thick] (3b);
\path[->] (3b)  edge  [thick] (1b);
\end{tikzpicture}
\caption{MDP of Figure~\ref{f_1} after applying policy $\pi$ defined in (\ref{e_pi})}
\label{f_2}
\end{figure}
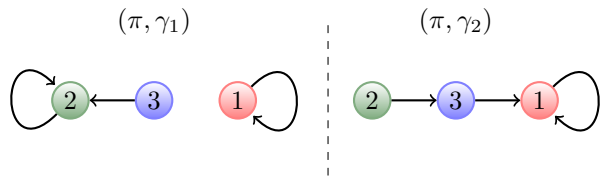

Next, we describe the methodology for deriving the Markov chain model from a MDP. A policy, $\pi$, is what determines which (low-level) control action to take at each step. It is the high-level control. Optimal and robust policies can be obtained using dynamic programming tools~\cite{saglam_robust_2014,bellman_markovian_1957}. Let's say, we decide to use policy
\begin{equation}
\label{e_pi}
\pi(x[n],\gamma[n])=\begin{cases}
\zeta_2 \text{ if } x[n]=x_2 \text{ and } \gamma[n]=\gamma_1
\\\zeta_1 \text{ if } x[n]=x_2 \text{ and } \gamma[n]=\gamma_2
\\\zeta_1 \text{ if } x[n]=x_3 \text{ and } \gamma[n]=\gamma_1
\\\zeta_2 \text{ if } x[n]=x_3 \text{ and } \gamma[n]=\gamma_2 .
\end{cases}
\end{equation}
Notice that we don't determine a control action for state $x_1$, since nothing can be done differently at halt state. When we use $\zeta[n]=\pi(x[n],\gamma[n])$, (\ref{e_first}) becomes
\begin{equation}
x[n+1]=h(x[n],\gamma[n],\pi(x[n],\gamma[n])),
\end{equation}
which is a function of the state and randomness only. The result is illustrated in Figure~\ref{f_2}.

\subsection{Distribution for $\gamma$}
The last step before obtaining an absorbing Markov chain is to assume a distribution for randomness. For Figure~\ref{f_2}, say $P(\gamma_2)=0.01$, i.e., with probability $0.01$, $\gamma[n]$ will be $\gamma_2$. We then obtain Figure~\ref{f_coin_absorbing}, which we already studied. More complicated systems will end up being surprisingly similar to~Figure~\ref{f_coin_absorbing}.

\section{Conclusion}

In this paper, we studied First Passage Time, which is a survival metric. After providing some motivational examples, we presented system-wide Mean First Passage Time (MFPT), which is calculated using the second largest eigenvalue of the stochastic transition matrix. We showed that for metastable systems, system-wide MFPT is an accurate indicator across a large set of states, including those frequently visited. We then introduced Mean First Passage Value (MFPV), which gives a more general value of interest, e.g., energy expenditure, distance, or time. We then provided bounds on First Passage Value (FPV) for a given confidence level. The last section of this paper was perhaps the most important one, because it showed how these tools explained can be used to low-level and high-level control a hybrid systems.

\section*{Acknowledgments}

This work was supported by the Institute for Collaborative Biotechnologies through grant W911NF-09-0001 from the U.S. Army Research Office. The content of the information does not necessarily reflect the position or the policy of the Government, and no official endorsement should be inferred.

We would also like to thank Beril Pisgin for drawing Figure~\ref{f_roadmap}.

\bibliographystyle{ieeetran}
\bibliography{saglam_references}

\begin{thebibliography}{10}
\providecommand{\url}[1]{#1}
\csname url@samestyle\endcsname
\providecommand{\newblock}{\relax}
\providecommand{\bibinfo}[2]{#2}
\providecommand{\BIBentrySTDinterwordspacing}{\spaceskip=0pt\relax}
\providecommand{\BIBentryALTinterwordstretchfactor}{4}
\providecommand{\BIBentryALTinterwordspacing}{\spaceskip=\fontdimen2\font plus
\BIBentryALTinterwordstretchfactor\fontdimen3\font minus
  \fontdimen4\font\relax}
\providecommand{\BIBforeignlanguage}[2]{{%
\expandafter\ifx\csname l@#1\endcsname\relax
\typeout{** WARNING: IEEEtran.bst: No hyphenation pattern has been}%
\typeout{** loaded for the language `#1'. Using the pattern for}%
\typeout{** the default language instead.}%
\else
\language=\csname l@#1\endcsname
\fi
#2}}
\providecommand{\BIBdecl}{\relax}
\BIBdecl

\bibitem{talkner_discrete_1987}
\BIBentryALTinterwordspacing
P.~Talkner, P.~Hanggi, E.~Freidkin, and D.~Trautmann,
  ``\BIBforeignlanguage{en}{Discrete dynamics and metastability: Mean first
  passage times and escape rates},'' \emph{\BIBforeignlanguage{en}{Journal of
  Statistical Physics}}, vol.~48, no. 1-2, pp. 231--254, Jul. 1987. [Online].
  Available: \url{http://link.springer.com/article/10.1007/BF01010408}
\BIBentrySTDinterwordspacing

\bibitem{hanggi_reaction-rate_1990}
\BIBentryALTinterwordspacing
P.~Hanggi, P.~Talkner, and M.~Borkovec, ``Reaction-rate theory: fifty years
  after kramers,'' \emph{Reviews of Modern Physics}, vol.~62, no.~2, pp.
  251--341, Apr. 1990. [Online]. Available:
  \url{http://link.aps.org/doi/10.1103/RevModPhys.62.251}
\BIBentrySTDinterwordspacing

\bibitem{muller_rates_1997}
\BIBentryALTinterwordspacing
R.~Muller, P.~Talkner, and P.~Reimann, ``Rates and mean first passage times,''
  \emph{Physica A: Statistical Mechanics and its Applications}, vol. 247, no.
  1–4, pp. 338--356, Dec. 1997. [Online]. Available:
  \url{http://www.sciencedirect.com/science/article/pii/S0378437197003907}
\BIBentrySTDinterwordspacing

\bibitem{kampen_stochastic_2011}
N.~Kampen, \emph{\BIBforeignlanguage{en}{Stochastic Processes in Physics and
  Chemistry}}.\hskip 1em plus 0.5em minus 0.4em\relax Elsevier, Aug. 2011.

\bibitem{larsen_like-charge_1997}
\BIBentryALTinterwordspacing
A.~E. Larsen and D.~G. Grier, ``\BIBforeignlanguage{en}{Like-charge attractions
  in metastable colloidal crystallites},''
  \emph{\BIBforeignlanguage{en}{Nature}}, vol. 385, no. 6613, pp. 230--233,
  Jan. 1997. [Online]. Available:
  \url{http://www.nature.com/nature/journal/v385/n6613/abs/385230a0.html}
\BIBentrySTDinterwordspacing

\bibitem{veendrick_behaviour_1980}
H.~J. Veendrick, ``The behaviour of flip-flops used as synchronizers and
  prediction of their failure rate,'' \emph{{IEEE} Journal of Solid-State
  Circuits}, vol.~15, no.~2, pp. 169--176, Apr. 1980.

\bibitem{fingelkurts_making_2004}
A.~A. Fingelkurts and A.~A. Fingelkurts, ``\BIBforeignlanguage{eng}{Making
  complexity simpler: multivariability and metastability in the brain},''
  \emph{\BIBforeignlanguage{eng}{The International Journal of Neuroscience}},
  vol. 114, no.~7, pp. 843--862, Jul. 2004.

\bibitem{byl_metastable_2009}
\BIBentryALTinterwordspacing
K.~Byl and R.~Tedrake, ``\BIBforeignlanguage{en}{Metastable walking
  machines},'' \emph{\BIBforeignlanguage{en}{The International Journal of
  Robotics Research}}, vol.~28, no.~8, pp. 1040--1064, Aug. 2009. [Online].
  Available: \url{http://ijr.sagepub.com/cgi/doi/10.1177/0278364909340446}
\BIBentrySTDinterwordspacing

\bibitem{benallegue_metastability_2013}
M.~Benallegue and J.-P. Laumond, ``Metastability for high-dimensional walking
  systems on stochastically rough terrain.'' in \emph{Robotics: Science and
  Systems}.\hskip 1em plus 0.5em minus 0.4em\relax Citeseer, 2013.

\bibitem{saglam_robust_2014}
C.~O. Saglam and K.~Byl, ``Robust policies via meshing for metastable rough
  terrain walking,'' in \emph{Proceedings of Robotics: Science and Systems},
  Berkeley, {USA}, 2014.

\bibitem{saglam_quantifying_2014}
------, ``Quantifying the trade-offs between stability versus energy use for
  underactuated biped walking,'' in \emph{{IEEE}/{RSJ} International Conference
  on Intelligent Robots and Systems ({IROS})}, 2014.

\bibitem{ahn_global_2013}
H.~J. Ahn and B.~Hassibi, ``Global dynamics of epidemic spread over complex
  networks,'' in \emph{2013 {IEEE} 52nd Annual Conference on Decision and
  Control ({CDC})}, Dec. 2013, pp. 4579--4585.

\bibitem{saglam_robust_2015}
C.~O. Saglam and K.~Byl, ``Robust policies via meshing for metastable
  walking,'' in \emph{International Journal Robotics Research (IJRR)}, 2015,
  invited and submitted.

\bibitem{saglam_lyapunov-based_2014}
C.~O. Saglam, A.~Teel, and K.~Byl, ``Lyapunov-based versus {P}oincare map
  analysis of the rimless wheel,'' in \emph{{IEEE} Conference on Decision and
  Control ({CDC})}, Dec. 2014, accepted for publication.

\bibitem{matthews_number_1995}
\BIBentryALTinterwordspacing
K.~Matthews, ``Number theory web,'' 1995. [Online]. Available:
  \url{http://www.numbertheory.org/courses/MP274/markov.pdf}
\BIBentrySTDinterwordspacing

\bibitem{meyer_matrix_2000}
C.~D. Meyer, \emph{\BIBforeignlanguage{en}{Matrix Analysis and Applied Linear
  Algebra}}.\hskip 1em plus 0.5em minus 0.4em\relax {SIAM}, Jun. 2000.

\bibitem{saglam_metastable_2014}
C.~O. Saglam and K.~Byl, ``Metastable markov chains,'' in \emph{{IEEE}
  Conference on Decision and Control ({CDC})}, Dec. 2014, accepted for
  publication.

\bibitem{goebel_hybrid_2012}
R.~Goebel, R.~G. Sanfelice, and A.~R. Teel,
  \emph{\BIBforeignlanguage{en}{Hybrid Dynamical Systems: Modeling, Stability,
  and Robustness}}.\hskip 1em plus 0.5em minus 0.4em\relax Princeton University
  Press, 2012.

\bibitem{bellman_markovian_1957}
\BIBentryALTinterwordspacing
R.~Bellman, ``A markovian decision process,'' \emph{Indiana University
  Mathematics Journal}, vol.~6, no.~4, pp. 679--684, 1957. [Online]. Available:
  \url{http://www.iumj.indiana.edu/IUMJ/FULLTEXT/1957/6/56038}
\BIBentrySTDinterwordspacing

\end{thebibliography}

\end{document}